\documentstyle[12pt,preprint]{aastex}

\def\ni{\noindent}

\def\lta{\lower2pt\hbox{$\buildrel {\scriptstyle <} 
   \over {\scriptstyle\sim}$}}
\def\gta{\lower2pt\hbox{$\buildrel {\scriptstyle >} 
   \over {\scriptstyle\sim}$}}

\newcount\eqnumber
\def\chaphead{}

\def\new{\hbox{\chaphead\the\eqnumber}\global\advance\eqnumber by 1}
\def\first{\hbox{(\chaphead\the\eqnumber{a}}\global\advance\eqnumber by 1}
\def\last{\advance\eqnumber by -1 \hbox{\chaphead\the\eqnumber}\advance
     \eqnumber by 1}
\def\eq#1{\advance\eqnumber by -#1 equation (\chaphead\the\eqnumber
     \advance\eqnumber by #1}
\def\eqnam#1{\xdef#1{\chaphead\the\eqnumber}}

\begin{document}

\title{The Evolution of a Structured Relativistic Jet and GRB Afterglow Light-Curves}
\author{Pawan Kumar}
\affil{Astronomy Department, University of Texas, Austin, TX 78731}
\author{Jonathan Granot}
\affil{Institute for Advanced Study, Princeton, NJ 08540}
\email{pk@astro.as.utexas.edu, granot@ias.edu}

\begin{abstract}

We carry out a numerical hydrodynamical modeling for the evolution 
of a relativistic collimated outflow, as it interacts with the 
surrounding medium, and calculate the light-curve resulting from synchrotron
emission of the shocked fluid. The hydrodynamic equations are reduced to 
1-D by assuming axial symmetry and integrating over the radial 
profile of the flow, thus considerably reducing the computation time.
We present results for a number of different initial jet structures, 
including several different power-laws and a Gaussian profile for 
the dependence of the energy per unit solid angle, $\epsilon$, and 
the Lorentz factor, $\Gamma$, on the angle from the jet symmetry axis.
Our choice of parameters for the various calculations is motivated by
the current knowledge of relativistic outflows from gamma-ray bursts and the 
observed afterglow light-curves. Comparison of the light curves for different
jet profiles with GRB afterglow observations provides constraints on the
jet structure. One of the main results we find is that the transverse fluid 
velocity in the comoving frame ($v_t$) and the speed of sideways expansion, 
for smooth jet profiles, is typically much smaller than the speed of sound 
($c_s$) throughout much of the evolution of the jet; $v_t$ approaches $c_s$ 
when  $\Gamma$ along the jet axis becomes of order a few (for large 
angular gradient of $\epsilon$, $v_t\sim c_s$ while $\Gamma$ is still 
large). This result suggests that the dynamics of relativistic structured 
jets may be reasonably described by a simple analytic model where 
$\epsilon$ is independent of time, as long as $\Gamma$ along the jet-axis 
is larger than a few.
\end{abstract}

 
\section{Introduction}

The great advance in our understanding of Gamma-Ray Bursts (GRBs) in the 
last five years has largely resulted from the observation and modeling of 
afterglow radiation -- emission observed for days to months after the 
end of a GRB, in the X-ray, optical and radio bands. 
The basic procedure for obtaining information about the explosion,
such as the energy release, opening angle of the emergent jet, the 
density of the medium in the immediate vicinity of the GRB etc., is by 
comparing the observed afterglow light-curve with the theoretically 
calculated flux (Wijers \& Galama 1999; Granot, Piran \& Sari 1999b; 
Chevalier \& Li 2000; Panaitescu \& Kumar 2001a,b, 2002).
Most works on GRB jets assume a homogeneous (or `top hat') jet, where all
the hydrodynamic quantities of the jet, such as its Lorentz factor and 
energy density, are the same within some finite, well defined, opening angle 
around the jet axis, and drop to zero at larger angles.

A comparison of theoretically calculated light-curves, under several 
simplifying assumptions described below (and assuming a `top hat' jet), 
with observed light-curves in X-ray, optical, and radio bands for 8 GRBs, 
has led to a number of interesting results (Panaitescu \& Kumar 2001b). 
Perhaps the most remarkable discovery is that the kinetic energy in the 
relativistic outflow is nearly the same, within a factor of 5, for the set 
of eight GRBs. A similar result has been obtained by Piran et al. (2001) 
through a method which requires fewer assumptions. Frail et al. (2001) 
have also found that the energy radiated in GRBs does not vary much from one 
burst to another. The opening angle for GRB jets is found to be in the range 
of 2--20 degrees, and the density of the external medium in the vicinity of 
GRBs is estimated to be between 10$^{-3}$ and 30 cm$^{-3}$.
Moreover, there is no firm evidence for the density to vary as inverse squared 
distance in all but one case (Price et al. 2002; Panaitescu \& Kumar, 2002), 
which is surprising in light of the currently popular model for 
GRBs -- the collapsar model.

The possibility that GRB jets can display an angular structure, i.e. that the 
Lorentz factor, $\gamma$, and energy per unit solid angle, $\epsilon$, in the 
GRB outflow can vary smoothly as power laws in the angle $\theta$ from the 
jet axis, was proposed by M\'esz\'aros, Rees \& Wijers (1998). Recently, 
in view of the evidence described above for a roughly constant energy in 
the gamma-ray emission and in the kinetic energy of the afterglow shock, 
it has been suggested that GRB jets might have a universal structure, 
and the differences in the observed properties of GRBs and their afterglows
arise due to different viewing angles, $\theta_{\rm obs}$, w.r.t the jet axis
(Lipunov, Postnov \&  Prokhorov  2001; Rossi, Lazzati \& Rees 2002; 
Zhang \& M\'esz\'aros 2002). 
In this interpretation, the jet break in the light curve occurs when 
the Lorentz factor along the line of sight, $\gamma(\theta_{\rm obs})$, 
drops to $\sim\theta_{\rm obs}^{-1}$, so that the jet break time, $t_{j}$, 
is determined by the viewing angle,  $\theta_{\rm obs}$, rather than by the 
opening angle of the `top hat' jet, as in the conventional interpretation.

The calculation of light-curves from a shock-heated, collimated, relativistic 
outflow has been carried out by a few research groups (Rhoads 1999; 
Panaitescu \& M\'esz\'aros 1999; Kumar \& Panaitescu 2000; Moderski, Sikora 
\& Bulik 2000; Granot et al. 2002). However, most of the works to date have 
been based on a simplified model for the jet dynamics and on a number of ad-hoc 
assumptions. All the above works assume a `top hat' jet, and furthermore, 
most of them model the dynamics of the jet at times much greater than 
the deceleration time
as uniform expansion at the sound speed or the speed of light (in the local 
rest frame of the shocked fluid)-- the results are nearly the same for 
both of these cases.
Similar simplifications were made in the recent work on a universal 
structured jet (Rossi, Lazzati \& Rees 2002). An exception to this, is the 
work of Granot et al. (2001), where the dynamics of an initial `top hat' 
jet were calculated using a hydrodynamic simulation, and the resulting 
light curves were calculated numerically.
However, such hydrodynamic simulations are very time consuming, and difficult 
to apply to a structured jet, so that there is currently no rigorous treatment
of the hydrodynamic evolution of a structured jet. In this paper we
develop such a rigorous treatment for the dynamics of structured jets, which 
at the same time is not very time consuming and may become practical to 
include in fits to afterglow observations.

Another simplification made in previous works (including all the works
mentioned above), and in the lack of a better alternative, is also made in 
this work, is that the strength of the magnetic field and the 
energy in the electrons are determined by assuming that 
the energy densities of the magnetic field and of the electrons are
constant fractions of the internal energy density of the shocked fluid. 
It is unclear how some of the simplifying assumptions in the afterglow
light-curve modeling effect the overall burst parameters and properties 
we have inferred as described above. 

Some progress has been made recently toward understanding the generation of 
magnetic fields in relativistic collisionless shocks: the numerical simulations 
of Medvedev (2002) show that 
magnetic fields generated behind collisionless relativistic shocks via the 
Weibel instability (Medvedev \& Loeb 1999) do not decay to very low values 
within a short distance behind the shock, as was previously thought 
(Gruzinov 1999, 2001), but rather approach a finite value in the bulk of the
 shocked fluid behind the shock, which might be compatible with the values 
inferred from afterglow observations.  Moreover,
the modeling of GRB afterglow light-curves indicates that the energy
fraction in electrons is close to equipartition (Panaitescu \& Kumar, 2001b),
hence the parametrization of electron energy does not appear to be a 
serious drawback for current models. 
Thus, at present, one of the biggest uncertainties in the afterglow modeling
is the assumption of a uniform jet and the simplified jet
dynamics. The purpose of this paper is to remedy this situation
and develop a much more realistic model for GRB jets. 
Fitting afterglow observations with light-curves that are calculated using
a realistic jet model \& dynamics may both constrain the structure of
GRB jets (the initial distribution of the Lorentz factor and energy per 
unit solid angle as a function of the angle from the jet axis), and provide 
more accurate estimates for the physical parameters, which include the 
external density profile and the parameters describing the micro-physics of 
relativistic collissionless shocks.

In the next section (\S \ref{hydro}) we discuss the evolution of structured 
jets. In \S \ref{dynamics} we describe our hydrodynamical scheme where we 
begin from the full hydrodynamic equations, assume axial symmetry and integrate 
over the radial structure, thus reducing the problem to a set of 
one dimensional partial differential equations that are solved numerically.
The initial and boundary conditions are outlined in \S \ref{IC}, while
results for some physically interesting cases are shown in \S \ref{NR}.
In \S \ref{LC} we describe the light-curve calculation and compare 
the results of hydro simulations with a simplified model. The main results 
are summarized in \S \ref{conc}.

\medskip
\section{\Large Jet modeling}
\label{hydro}
\medskip

We begin with a brief description of the uniform jet model, 
and then we describe in some detail the evolution 
of a more realistic, structured, jet and the afterglow light-curves
resulting from emission by the shock heated medium swept up by the jet.

Most calculations of GRB light-curves have assumed that the properties of 
the relativistic outflow do not vary across the jet, and that the jet
dynamics is described by a uniform lateral expansion in the comoving frame,
at close to the speed of sound, $c_s$, 
which for a hot relativistic plasma is $3^{-1/2}$ times the speed of light,
$c$. These assumptions drastically simplify the calculation of the 
evolution of the jet opening angle, $\theta_j$, with time: 
the increase in the lateral size of the jet 
in comoving time $\delta t_{co}$ is 
$c_s\delta t_{co}$, and so the change to its angular size is $\delta\theta_j = 
c_s \delta t_{co}/r=(c_s/c)\delta r/(r\gamma)$, or, 

$$ {d\theta_j\over dr} = {c_s\over c\gamma r}\ . $$

\noindent This equation, together with the energy conservation equation, 
describe the dynamics of a uniform relativistic disk or a jet. The
implication of this equation is that the jet opening angle 
$\theta_j$ starts to increase when $\gamma$ drops below $\theta_j^{-1}$,
and from that time onward the jet opening angle is roughly $\gamma^{-1}$. 
A detailed discussion of the uniform jet dynamics and lightcurve calculation 
can be found in a number of papers (e.g. Rhoads 1999; Panaitescu \& 
M\'esz\'aros 1999; Sari et al. 1999; Kumar \& Panaitescu 2000). 
Such a uniform jet with sharp well defined edges shall be referred
to as a `top hat' jet.

However,  2D hydrodynamical simulations of the evolution of
a jet that is initially uniform within some finite opening angle 
(Granot et al. 2001) have shown that the lateral expansion of the jet is
smaller than that predicted by the simple models described above. 
This suggests that the assumption of lateral expansion at close 
to the sound speed in the comoving frame, that is made in most simple
jet models, is not valid. Nevertheless, the light curves calculated 
from these simulations show a sharp jet break in the light curves, 
similar to that seen in most afterglow observations, around the time
$\gamma$ drops to $\theta_j^{-1}$.

\subsection{\Large Dynamics of Structured Relativistic jets}
\label{dynamics}

Clearly, it is unrealistic to assume that the outflow from GRB explosions 
will be uniform within some finite opening angle, outside of which the Lorentz 
factor, $\gamma$, and energy per unit solid angle, $\epsilon$, decrease 
very sharply (i.e. a `top hat' jet). A more realistic situation is that 
the Lorentz factor (LF), the energy density etc. are smooth functions of 
the angle, $\theta$, from the jet axis, and possibly also of the distance,
$r$, from the central source. The causality consideration suggests that the 
outflow is unlikely to be uniform over large angles, 
and moreover it provides a limit 
on how rapidly initial inhomogeneities can be smoothed out. Let the LF of the 
shell after elapsed time $t$ since the explosion, measured in the lab frame, 
be $\gamma(t)$. The comoving time corresponding to this is $\sim t/\gamma$, 
and the distance traversed by sound waves during this interval is 
$c_s t/\gamma\sim c t/3^{1/2}\gamma$. Therefore, the angular size of a 
causally connected region is
$\sim 1/3^{1/2}\gamma$, and inhomogeneities on an angular scale of 
$\theta_{ih}>\gamma^{-1}$, if present initially, will persist; the 
inhomogeneities can be smoothed out only when the LF has dropped below 
$\theta_{ih}^{-1}$. As an example, the large angular scale inhomogeneities 
for a jet of opening angle $5^o$ start to decrease only when the bulk LF has 
dropped below $\sim 10$, or roughly one day after the explosion (as seen by 
the observer). It should also be noted that if one were to start with a 
uniform jet, or a top-hat profile for the LF or $\epsilon$, the large 
gradient at the edge will decrease with time and the jet will develop 
angular structure (e.g. Granot et al. 2001).

The remainder of this section is devoted to the solution of the relativistic
hydrodynamic equations to describe the evolution of jets from GRBs.
The starting point is the relativistic fluid equations 
(e.g. Landau \& Lifshitz, 1959):
\begin{equation}
\label{T_mu_nu}
T^{\mu\nu}_{\ \ \,;\nu} = 0\quad , \quad 
T^{\mu\nu} = w u^\mu u^\nu + p g^{\mu\nu} \ ,
\end{equation}
where $T^{\mu\nu}$ is the energy-momentum tensor for an 
ideal fluid, $u^\mu$ is the 4-velocity of the fluid, 
$g^{\mu\nu}$ is the metric tensor, $p$ is the pressure and $w=\rho+e+p$ 
is the proper enthalpy density, where $\rho$ and $e$ are the proper rest 
mass density and internal energy density, respectively, and $c=1$ in our units.
We use a spherical coordinate system and assume the flow
possesses axial symmetry about the z-axis, i.e. $u^{\phi},\partial/\partial\phi
=0$. Under these assumptions the $t,r$ and $\theta$ components of equation 
(\ref{T_mu_nu}) are
\vspace{0.2cm}
\begin{equation}
\label{t}
{\partial\over\partial t}(w\Gamma^2-p)
+{1\over r^2}{\partial\over\partial r}(r^2 w\Gamma^2 v_r)
+{1\over r\sin\theta}{\partial\over\partial\theta}
(\sin\theta\, w\Gamma^2 v_\theta) = 0 \ ,
\end{equation}
\vspace{0.2cm}
\begin{equation}
\label{r}
{\partial\over\partial t}(w\Gamma^2 v_r)
+{1\over r^2}{\partial\over\partial r}(r^2 w\Gamma^2 v_r^2)
+{1\over r\sin\theta}{\partial\over\partial\theta}
(\sin\theta\, w\Gamma^2 v_r v_\theta)
+{\partial p\over\partial r}
-{w\Gamma^2 v_\theta^2\over r} = 0 \ ,
\end{equation}
\vspace{0.2cm}
\begin{equation}
\label{theta}
{\partial\over\partial t}(w\Gamma^2 v_\theta)
+{1\over r^2}{\partial\over\partial r}(r^2 w\Gamma^2 v_r v_\theta)
+{1\over r\sin\theta}{\partial\over\partial\theta}
(\sin\theta\, w\Gamma^2 v_\theta^2)
+{1\over r}{\partial p\over\partial\theta}
+{w\Gamma^2 v_r v_\theta\over r} = 0 \ ,
\end{equation}
\vspace{-0.2cm}\\
where $v_r$ and $v_\theta$ are the $r$ and $\theta$ components of the
fluid velocity, and $\Gamma=(1-v_r^2-v_\theta^2)^{-1/2}$ is the Lorentz
factor of the fluid. Assuming that pair production has a negligible effect on
the rest mass density, baryon number conservation implies
\vspace{0.2cm}
\begin{equation}
\label{mass}
(\rho u^\mu)_{;\mu}={\partial\over\partial t}(\rho\Gamma)
    +{1\over r^2}{\partial\over\partial r}(r^2\rho\Gamma v_r)
    +{1\over r\sin\theta}{\partial\over\partial\theta}
      (\sin\theta\, \rho\Gamma v_\theta)=0\ .
\end{equation}
\vspace{-0.2cm}\\
We assume an equation of state
\begin{equation}
\label{eos}
 p=(\hat{\gamma}-1)e={\hat{\gamma}-1\over\hat{\gamma}}(w-\rho)\ \quad\quad
 {\rm with} \quad\quad \hat\gamma = {4\Gamma + 1\over3\Gamma}\ .
\end{equation}

Equations (\ref{t})--(\ref{eos}) can be solved together to determine the 
structure and evolution of the outflow from GRBs. The computation time, for 
a 1 GHz clock speed computer, and for a modest resolution in r \& $\theta$ 
coordinate of 100x1000 (in order to keep the error small in finite 
difference schemes) and 5000 time steps, is expected to take of order 
several hours to complete one run for one set of initial conditions; for
comparison the 2-D relativistic jet hydrodynamics calculation 
of Miller and Hughes, reported in Granot et al. (2001), took several
hours to days of computation time, for low to medium resolution runs, to 
follow the evolution for $\gtrsim 10$ observer days, while an even longer 
computational time was required for the higher resolution runs.
The successful modeling of light-curves of a single GRB to determine various 
parameters requires several thousand runs, and thus the computation time to
model one GRB, using a 2-D code, is currently estimated to range between months 
to years. Using many processors simultaneously can help reduce the actual
overall time required, but at any rate, this requires a great computational 
effort.

The computation time can be drastically decreased by reducing the
problem to a 1-D system, by integrating
out the radial dependence for all of the relevant variables, over the width 
of the outflow plus the swept up material. The physical motivation for 
this is that jets in GRBs are in fact thin 
shells.\footnote{At a distance $r$ from the center of the explosion, the 
laboratory frame radial thickness of the ejecta plus the swept-up
shock heated material moving with LF $\gamma$ is $\sim r/4\gamma^2$,
whereas its transverse dimension is $r\theta_j$. Therefore the 
geometric shape of the system is that of a thin
disk as long as $\theta_j\gg 1/4\gamma^2$.}
This procedure reduces the
computing time drastically without introducing a significant loss of
information as far as the emergent synchrotron emission is concerned;
we find that the lightcurves from a relativistic spherical shock which has 
radial structure described by the self-similar Blandford-McKee (1976) solution 
is almost the same as in a model where the radial dependences have been
integrated out and the shell thickness is taken to be zero
(see Figure 5 of Granot, Piran \& Sari 1999a).

The shock front is a two dimensional surface described
by $r=R(\theta,t)$. The shocked fluid is concentrated in a thin shell of
thickness $\Delta R\sim R/4\Gamma^2\ll R$ for a relativistic
flow\footnote{In fact, even as the flow becomes non-relativistic, we still
expect $\Delta R/R\lesssim 0.1$, as in the Sedov-Taylor self-similar 
solution, so that the thin shell approximation should still be reasonable.},
and therefore it makes sense to integrate all the dependent variables,
such as $p$, $w$, $\Gamma$, etc., over the width of the shell in the radial 
direction. We define quantities averaged over $r$, at a fixed $\theta$ 
and lab frame time $t$, as follows:

\begin{equation}
\label{def_a}
{\Pi}=\int_0^R dr\,r^2\, p\quad , \quad  \Pi\,{\overline\Gamma^2}=
    \int_0^R dr\, r^2\, p \Gamma^2\quad , \quad {\overline\Gamma}\,\overline
     u_\theta \Pi=\int_0^R dr\, r^2\, p\Gamma u_\theta \ ,
\end{equation}

\begin{equation}
\label{def_c}
 \Pi\,{\overline\Gamma}\,\overline u_r =\int_0^R dr\, r^2\,  p\Gamma u_r\quad , \quad
   \chi_1 \Pi\,\overline u_r\overline u_\theta=\int_0^R dr\,
   r^2\, p\, u_r u_\theta \quad , \quad
   \chi_2\,\overline u_\theta^{\,2}\,\Pi=\int_0^R dr\, r^2\, p u_\theta^2 \ ,
\end{equation}

\begin{equation}
\label{def_mu}
\mu_s\equiv{dM_s\over d\Omega}=\int_0^R dr\,r^2
\,\rho\Gamma\quad ,
\quad \mu_0\equiv{dM_0\over d\Omega}=\int_0^R dr\,r^2\,
\rho_0\Gamma \ ,
\end{equation}
where $\mu_s$ and $\mu_0$ are the rest mass per unit solid angle of swept-up 
material and of the original ejecta, respectively, and $\chi_1$ \& 
$\chi_2$ are dimensionless correlation coefficients, of order unity 
magnitude, which are taken to be independent of time. 
Integration of equation (\ref{t}) times $r^2$, over the radial interval 
corresponding to the width of the shell, when the shell is located at $R(t)$, 
yields

\begin{equation}
\label{epsilon}
{\partial\over\partial t}\left[ \xi\overline\Gamma^2 - \Pi
   \right]  + {1\over\sin\theta}
{\partial\over\partial\theta} \left[
{\sin\theta\, u_\theta\overline\Gamma \xi\over R}\right]
= \rho_{\rm ext}(R)R^2{\partial R\over\partial t}\ ,
\end{equation}
where
\begin{equation}
\label{A}
\xi\equiv {\mu_0 + \mu_s\over \overline\Gamma} + {\hat\gamma\over 
   \hat\gamma  - 1} \Pi \ .
\end{equation}

\ni An integration of equations (\ref{r}) and (\ref{theta}) over the width
of the shell gives

\begin{equation}
\label{epsil}
{\partial\over\partial t}\left[ \xi\overline\Gamma\overline u_r \right]
  + {1\over\sin\theta}
{\partial\over\partial\theta} \left[
{\sin\theta\, \chi_1 \overline u_\theta\overline u_r\xi\over R}\right]
 - {\chi_2 \xi \overline u^2_\theta\over R} - {2 \Pi\over R} = 0 \ ,
\end{equation}

\begin{equation}
\label{eps}
{\partial\over\partial t}\left[ \xi\overline\Gamma\overline u_\theta \right]
  + {1\over \sin\theta}
{\partial\over\partial\theta} \left[
{\sin\theta\, \chi_2 \overline u^2_\theta\xi\over R}\right]
 + { \xi \chi_1 \overline u_r \overline u_\theta\over R} + {1\over R}
   {\partial \Pi\over \partial \theta} = 0 \ ,
\end{equation}

\ni and the closure relation, given below, is obtained by integrating the 
equation $\Gamma^2 = 1 + u_r^2 + u_\theta^2$, times the pressure $p$, over 
the width of the shell

\begin{equation}
\label{close}
\overline\Gamma^2 - \chi_3\overline u_r^2 - \chi_2 \overline u_\theta^2 = 1\ .
\end{equation}
Where $\chi_3\approx 1$ is a constant factor; $\chi_3=1$ corresponds 
to the assumption that $1- \overline u_r/\overline\Gamma \ll 1$.

In deriving these equations it was assumed that the ejecta moves with the
shocked ISM, i.e. that the radial and the transverse components of the ejecta 
velocity are same as those of the swept up ISM. Under this assumption the mass
continuity equation for the ejecta and the shocked ISM are respectively

\begin{equation}
\label{massa}
{\partial \mu_0\over\partial t} + {1\over \sin\theta}
{\partial\over\partial\theta} \left[ {\sin\theta\, \overline u_\theta \mu_0
  \over R\,\overline\Gamma} \right] = 0,
\end{equation}

\begin{equation}
\label{massb}
{\partial\mu_s\over\partial t} + {1\over \sin\theta}
{\partial\over\partial\theta} \left[ {\sin\theta\, \overline u_\theta \mu_s
\over R\,\overline\Gamma} \right] = \rho_{\rm ext}(R) R^2 
   {\partial R\over\partial t}\ ,
\end{equation}

The velocity of the shock front is given by the shock jump conditions 
(Blandford \& McKee 1976):
\begin{equation}\label{BM}
v_{sh}=\sqrt{1-{\hat{\gamma}(2-\hat{\gamma})(\Gamma_{ps}-1)+2\over
(\Gamma_{ps}+1)\left[\hat{\gamma}(\Gamma_{ps}-1)+1\right]^2}}
\end{equation}
where $\Gamma_{ps}\approx \bar\Gamma$ is the post shock LF.
The shock jump conditions imply that in the rest frame of the fluid before the 
shock (which in our case is the lab frame), the direction of the velocity 
of the fluid just behind the shock is always perpendicular to the shock front.
In order to propagate the shock front in time, we assume that the average 
velocity of the shocked fluid is a good approximation for its value just behind 
the shock, and obtain
\begin{equation}\label{rad}
{\partial R\over\partial t}={v_{sh}\over\sqrt{1+(\bar{u}_\theta/\bar{u}_r)^2}}\ .
\end{equation}

Equations (\ref{epsilon})-(\ref{rad}) are solved numerically
with appropriate initial conditions (discussed below), to determine
the evolution of the jet. 

\subsection{\bf Initial \& boundary conditions}
\label{IC}

The initial conditions are chosen at a lab frame time $t_0$, after the internal 
shocks have ended, and before there is a significant deceleration due to the 
sweeping up of the external medium. We implement a number of different
initial conditions, one of which is that the initial energy (including
the rest mass energy) per unit 
solid angle, $\epsilon$, and the initial Lorentz factor (minus 1) are
power-law functions of $\theta$, outside of a core of opening angle 
$\theta_c$. The initial energy per unit solid angle, 
$\epsilon(\theta,t_0)$, and the LF for the power-law model are
\begin{equation}
\label{initial_con1}
\epsilon(\theta,t_0)
=\epsilon_0\Theta^{-a}\quad 
  ,\quad\Gamma(\theta,t_0)=1+(\Gamma_0-1)\Theta^{-b}\ ,
\end{equation}
where $\epsilon_0$ and $\Gamma_0$ are the initial energy per unit
solid angle and Lorentz factor at the jet axis, and
\begin{equation}\label{Theta}
\Theta\equiv\sqrt{1+\left({\theta\over\theta_c}\right)^2}\ .
\end{equation}

Another initial condition we explore is a Gaussian profile
for which $\epsilon(\theta,t_0)$ and $\Gamma(\theta,t_0)$ are
proportional to $\exp(-\theta^2/2\theta_c^2)$. The Gaussian
jet profile was mentioned in Zhang \& Meszaros (2001), however they
did not calculate the jet dynamics or lightcurve for this case. 

Equation (\ref{initial_con1}), or its counterpart for the Gaussian case,
are applied only as long as $\Gamma(\theta,t_0)>1.1$. At larger angles,
we assume a uniform outflow, the parameters of which are set by the
continuity condition.

We assume that the velocity is initially purely in the radial direction,
i.e.,
\begin{equation}\label{initial_vel}
v_\theta(\theta,t_0)=0\quad ,\quad 
v_r(\theta,t_0)=\sqrt{1-\Gamma^{-2}(\theta,t_0)}\ ,
\end{equation}
while the initial radius is given by $R(\theta,t_0) = t_0\,v_r(\theta,t_0)$.

The angular derivative of all dependent variables except $\overline{u}_\theta$ 
is zero at the pole and the equator, whereas $\partial\overline{u}_\theta/
\partial\theta$ at the pole is determined by the assumption of axisymmetry, 
and at the equator by the reflection symmetry; $\overline{u}_\theta$
vanishes at the pole and at the equator.

\subsection{\bf Numerical Results}
\label{NR}

Equations (\ref{epsilon})-(\ref{rad}) are solved using the two step 
Lax-Wendroff 
scheme, to determine the evolution of the jet, for several different choices 
of initial conditions. The number of grid points in the angular 
directions is taken to be about 2000, and the time step is chosen to satisfy
the Courant condition. The numerical solution respects
global energy conservation to within 0.1\%.
We have looked into the dependence of the solution on the value of
the dimensionless correlation parameters $\chi_1$ and $\chi_2$,
and find consistent solutions for $\chi_1$  between 0.8 and 1, and
$\chi_2$ approximately between 0.9 and 1.2. Outside of this range of
parameters the code is unstable and the solution unphysical, and the energy 
conservation is not satisfied. Inside this range the solution is not
sensitive to the exact value of $\chi_1$ and $\chi_2$.

\begin{figure}
\hspace{2.7cm}
\includegraphics[width=9.8cm]{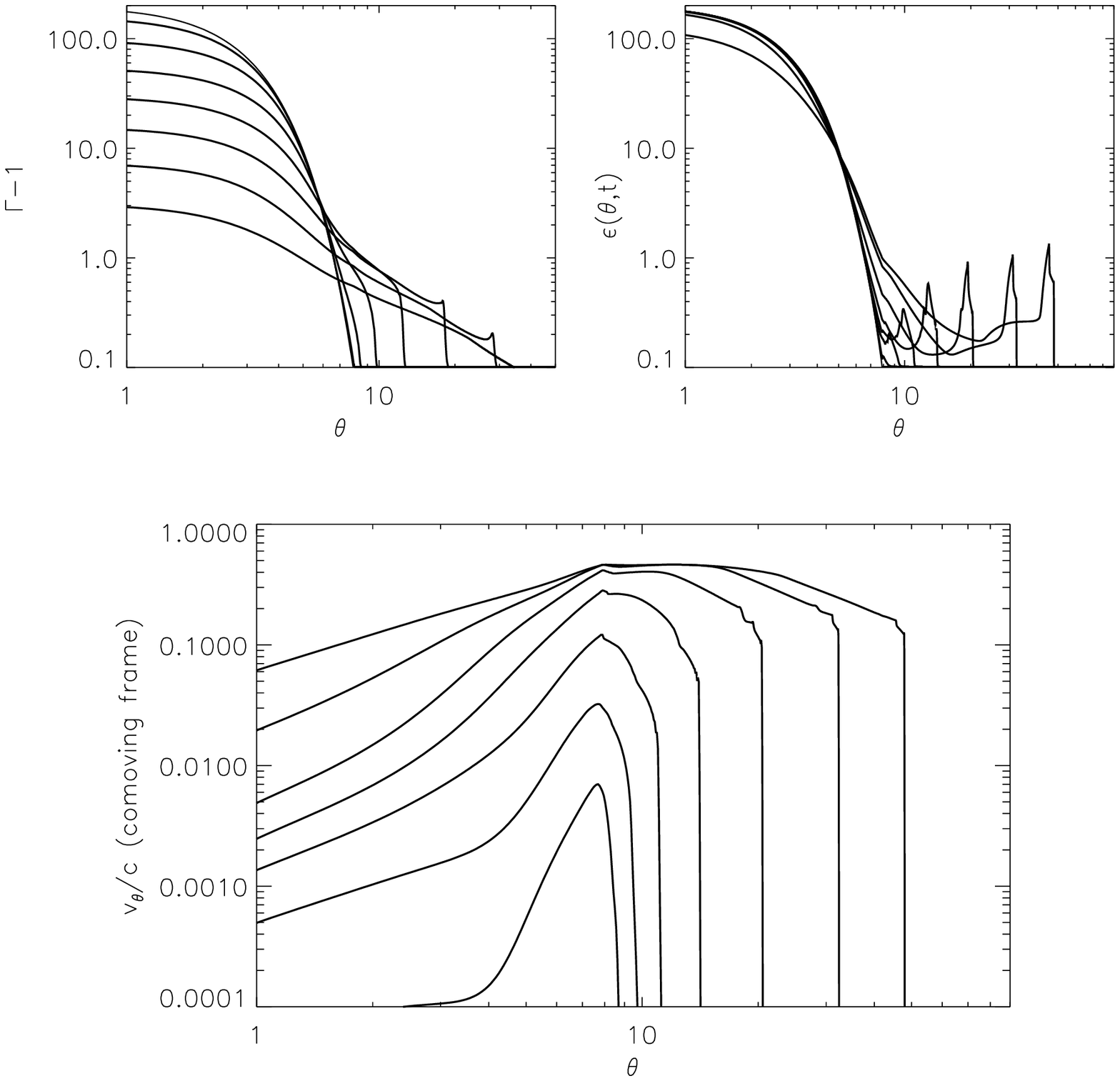}
\caption{\label{fig1a}The different panels show the evolution of the 
Lorentz factor, $\overline\Gamma(\theta,t)$, the energy per unit 
solid angle, $\epsilon(\theta,t)$, and the transverse velocity in 
the comoving frame, $v'_\theta/c = \overline u_\theta$,
for a jet with an initially Gaussian profile, i.e. $\overline\Gamma$ \&
$\epsilon$ proportional to $\exp(-\theta^2/2\theta_c^2)$ at the initial
time. The parameters are:  $\theta_c=0.035$ radian, $\epsilon(\theta=0,t_0)=
10^{53}/4\pi$ erg/sr, $\overline\Gamma(\theta=0,t_0)=200$, and the
density of the external medium is 10 particles per CC. At angles larger
than about 6.5$^o$ when $\overline\Gamma$ drops below 1.1, the LF and
$\epsilon$ are taken to be independent of $\theta$. Note that the energy
per unit solid angle does not change much with time except at large
angles where it was small initially, and we see it increase with time. 
The sideways expansion of the jet can be seen in the bottom panel, which 
shows the edge of the ``relativistic jet'', i.e. the vertical line where the
transverse velocity drops to zero, 
which is moving to larger angles with time; 
the jet edge, and its motion, can also be seen in the top left panel 
as a sharp drop in $\Gamma-1$. The sharp jump in the lateral velocity and 
energy density may be understood as a formation of a shock wave in the 
lateral direction.
The numerical scheme we use becomes unstable when $\overline\Gamma$ drops 
below 4 at $\theta=0$ or $v_\theta/c$ becomes larger than about 0.4; 
the LF near the edge of the jet is close to 1.01 when the code becomes 
unstable. It should be noted that the transverse velocity depends on the 
gradient of the
 energy density in the transverse direction and on the value of LF
 locally and not along the jet axis. And so our result for
 the transverse velocity -- that it remaines below the sound speed
 throughout much of the jet evolution -- is not compromised by the 
 numerical instability. The lab-frame-time increases monotonically from 
the curve with smallest $v_\theta/c$ to the curve with highest
$v_\theta/c$ (the lower panel), and the time increases from the
highest to the lowest curves in the top two panels.
}
\end{figure}

\begin{figure}
\includegraphics[width=15cm]{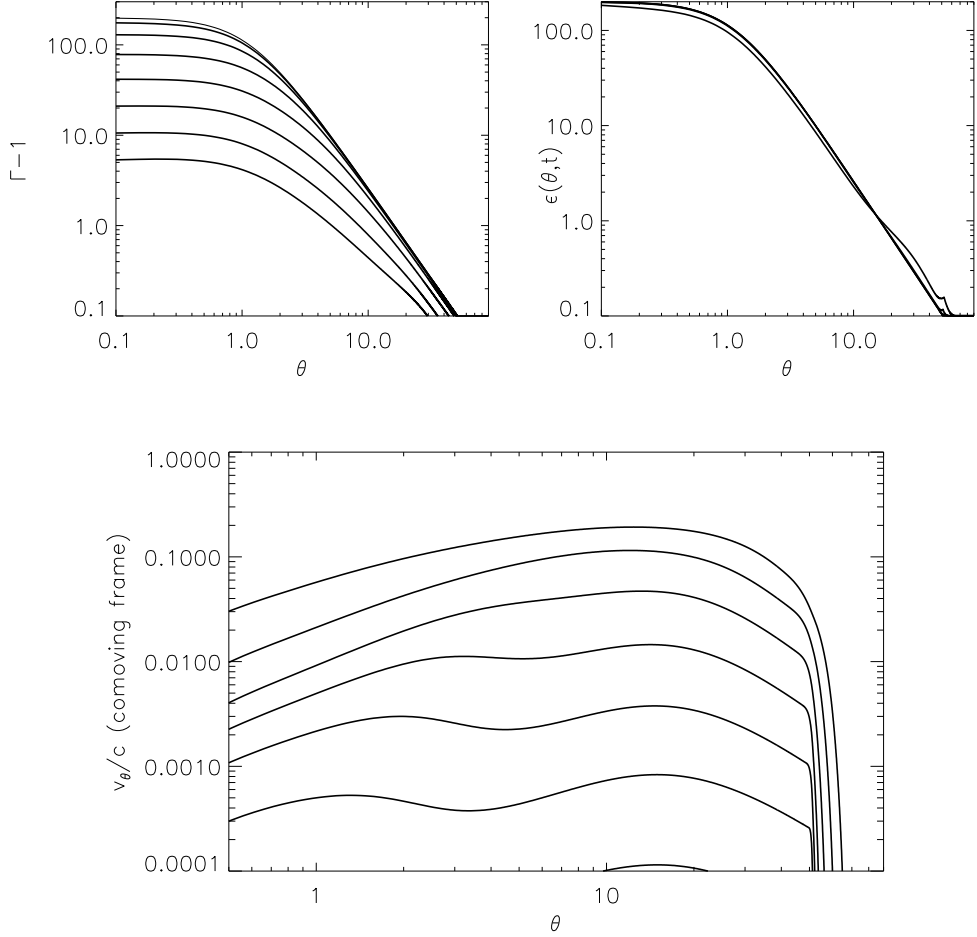}
\caption{\label{fig1b}The different panels show the evolution of the 
Lorentz factor $\overline\Gamma(\theta,t)$, the energy per unit 
solid angle, $\epsilon(\theta,t)$, and the transverse velocity in 
the comoving frame, $v'_\theta/c = \overline u_\theta$, 
for a jet with initial $\overline\Gamma=1 + 199/(1+\theta^2/\theta_c^2)$ and
$\epsilon=\epsilon_0(1+\theta^2/\theta_c^2)^{-1}$ with $\epsilon_0=
10^{53}/4\pi$ erg/sr, and $\theta_c=0.02$ radian; this initial profile
corresponds to $a=2$ \& $b=2$ in the notation of equation (\ref{initial_con1}).
The density of the external medium is 10 particles per CC. Note that the 
memory of the initial core angle, $\theta_c$, is not erased with time,
and in fact it remains unchanged until quite late times (see top left panel).
Moreover, the energy per unit solid angle does not change much with time 
either, except at large angles where it was initially small, and we see 
it increase slightly with time. The transverse velocity $v'_\theta$ (see lower
panel) remains quite small throughout much of the evolution of jet and
even at late stages is about a factor of 3 smaller than the sound speed;
the oscillations seen at early time in $v_\theta$ are almost
certainly unphysical and numerical in origin.
}
\end{figure}

\begin{figure}
\includegraphics[width=14cm]{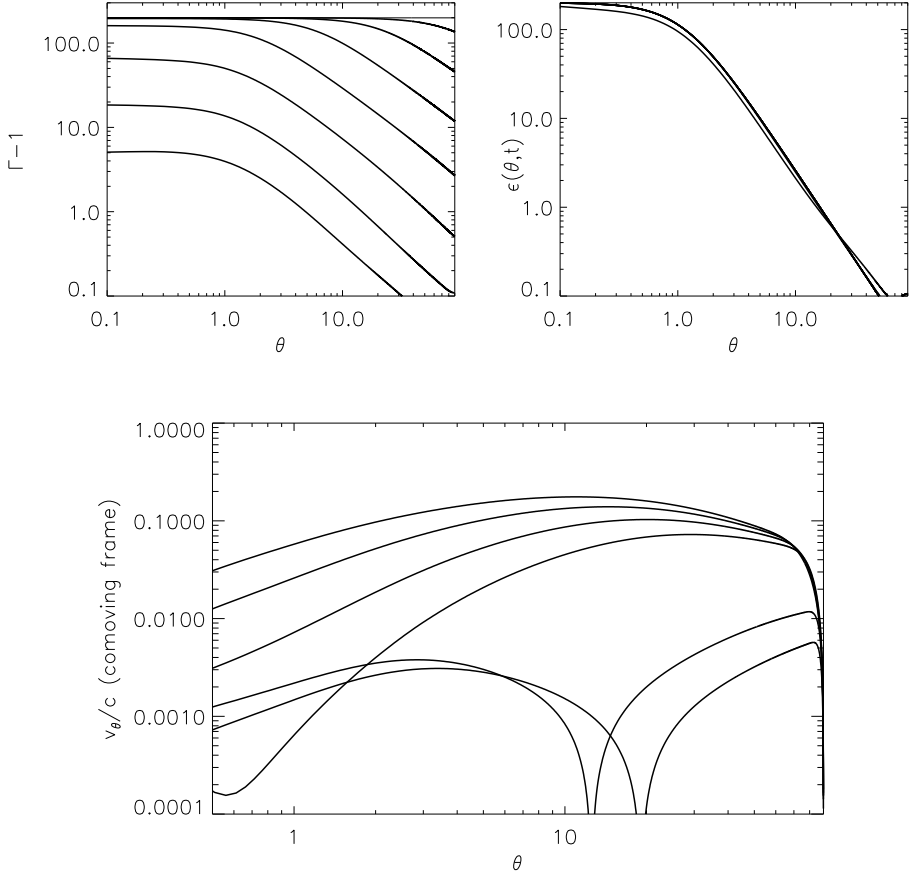}
\caption{\label{fig1c}Same as figure 2, expect that $a=2$ \& $b=0$ i.e. 
initially $\epsilon=\epsilon_0(1+\theta^2/\theta_c^2)^{-1}$ and 
$\overline\Gamma(\theta,t_0)=200$, with $\theta_c=0.02$ radian.
See caption for fig. 2 for other details. The minimum in $v_\theta$
at early times (the two lower curves on the bottom panel) is because
$v_\theta$ is changing sign from negative values at small angles to
positive values at larger angles and we have plotted $|v_\theta|$.
The reason for the negative velocity near the jet axis at early 
times is that in this model the deceleration time is largest at the
pole ($R_{\rm dec},\,t_{\rm dec}\propto\Theta^{-2/3}=
(1+\theta^2/\theta_c^2)^{-1/3}$); thus the pressure integrated over
the shell thickness is close-to-zero at the pole and non-zero at
large $\theta$ at times much less that the deceleration timescale
at the pole. In this case a pole-ward transverse flow ensues.
At $\theta>\theta_{\rm dec}(t)$ the $\epsilon\propto\Theta^{-2}$ profile 
dominates and induces a flow toward larger $\theta$.  
 At later time, much greater the deceleration time at the pole, the
pressure integrated over the shell thickness, is indeed largest
at the pole, as the referee mentioned, and the transverse velocity
becomes positive everywhere. For (a,b)=(2,1) model the transverse velocity
 is everywhere positive at all times as expected from this argument and
we have varified this numerically.
}
\end{figure}

\begin{figure}
\includegraphics[width=16cm]{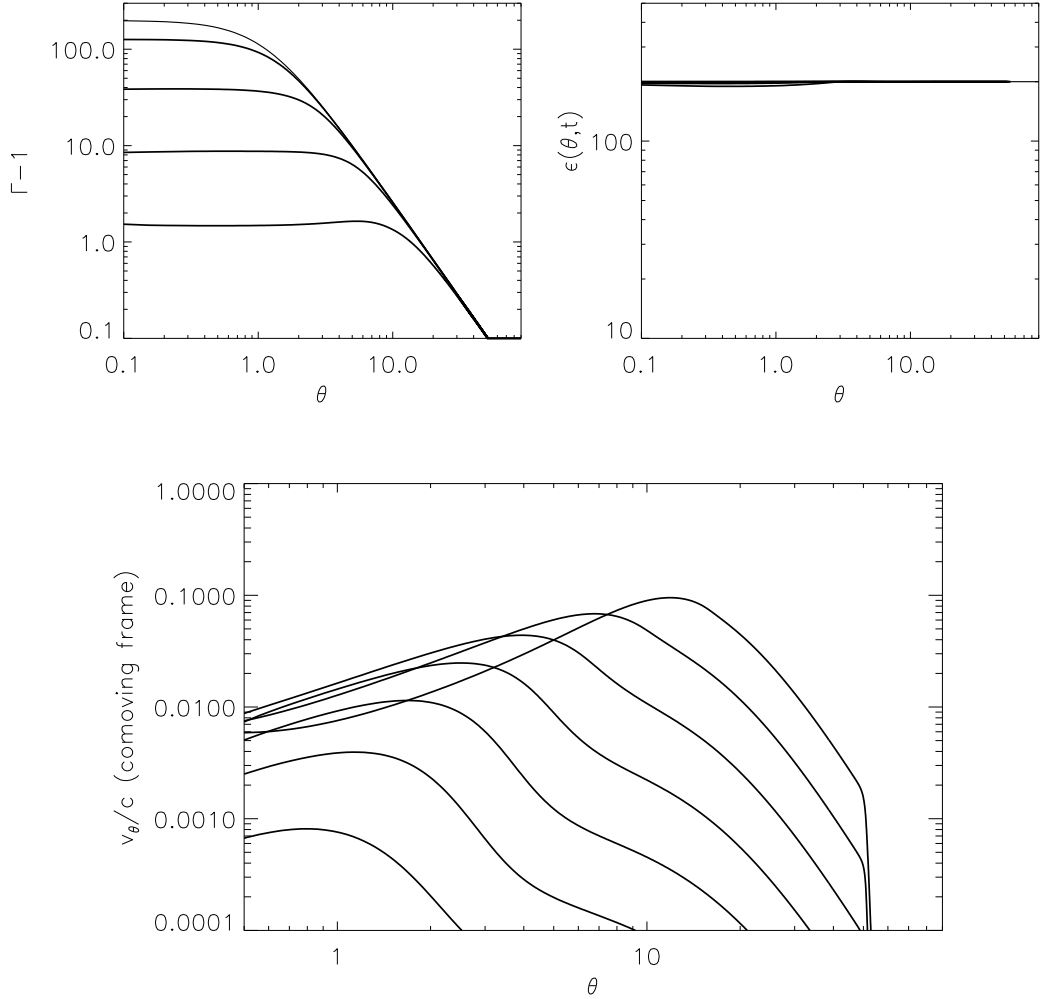}
\caption{\label{fig1d}Same as Figure \ref{fig1c}, expect that $a=0$ \& $b=2$, i.e. 
initially $\epsilon(\theta,t_0)=10^{53}/4\pi$ erg/sr, and $\overline\Gamma=1 + 
199/(1+\theta^2/\theta_c^2)$, with $\theta_c=0.02\;$rad.
See caption of Fig. \ref{fig1b} for other details. Note that the jet becomes
increasingly spherically symmetric with time in this case; 
$\Gamma$ is independent of $\theta$ up to an angle where shell deceleration
has occurred, as in fact we expect when $\epsilon$ is independent of angle.
}
\end{figure}

The evolution of $\overline{\Gamma}(\theta)$, $\overline{u}_\theta(\theta)$, 
and $\epsilon(\theta)$ are shown in figures \ref{fig1a}-\ref{fig1d} for 
$(a,b)=$ (0,2), (2,0), (2,2), and for a Gaussian jet. Note that the transverse 
velocity in the comoving frame of the shocked fluid, $v'_\theta=
\overline u_\theta$, is much less than the sound speed, $3^{-1/2}$, 
throughout much of the time,
and approaches the sound speed only when the jet Lorentz factor on the
axis has fallen to a value of order a few. Clearly, this result depends
on the gradient of the initial LF or $\epsilon$ at the initial time,
and therefore the transverse velocity is found to be largest for the
Gaussian case, which has the highest gradient of all the models we have
considered. 

The small value for $v'_\theta$ can be understood 
from equation (\ref{eps}). Ignoring the second order term in $\bar{u}_\theta$,
and noticing that the ``source term'' for $\bar{u}_\theta$ is the gradient of
$\Pi$, we find that $\overline u_\theta\sim 
(\overline\Gamma\delta\theta)^{-1}$, where $\delta\theta$ is the 
angular scale for the variation of $\Pi$ or the energy density.
Thus, we get an appreciable transverse velocity in the comoving frame
only when $\bar\Gamma\delta\theta\lesssim 1$.

Another way of deriving this result is from the shock jump conditions,
that imply that the velocity, ${\bf v}_{ps}$, of the fluid 
just behind the shock in the rest frame of the fluid before the shock 
(the lab frame in our case) is perpendicular to the shock front. 
This implies that the angle $\alpha$ between $\hat{v}_{ps}$ and $\hat{r}$ 
(i.e. $\hat{v}_{ps}\cdot\hat{r}=\cos\alpha$) satisfies
\begin{equation}\label{alpha}
\tan\alpha={v_\theta\over v_r}=-{1\over R}{\partial R\over\partial\theta}\ . 
\end{equation}
For a relativistic flow $v_\theta<\Gamma^{-1}\ll 1$ so that 
$v_\theta\ll v_r\approx 1$ and $\alpha\approx\tan\alpha\approx v_\theta$.
Thus we have $v_\theta\approx-\partial\ln R/\partial\theta$, and since
$R\approx(1-1/2\Gamma^2)t$, this implies 
$v_\theta\approx-\Gamma^{-3}\partial\Gamma/\partial\theta$,
or $v_\theta\sim1/(\Gamma^{2}\delta\theta)$ and 
$u_\theta=\Gamma v_\theta\sim 1/(\Gamma\delta\theta)$, where $\delta\theta$
is the angle over which $\Gamma$ varies appreciably. Therefore, this reproduces
the result of the previous paragraph, as it is easy to show that the angular scale 
for the variation of $\Pi$ and $\Gamma$ are similar. From the definition of $\Pi$ 
(equation \ref{def_a}), we have $\Pi\sim pR^2\Delta R\sim pR^3/4\Gamma^2$ and since 
the shock jump conditions imply $3p=e=4\Gamma^2\rho_{ext}(R)$, this gives
$\Pi\sim\rho_{ext}(R)R^3/3$, i.e. $\Pi$ is a power law in 
$R\approx(1-1/2\Gamma^2)t$, and therefore varies over the same angular scales as 
$\Gamma$.

The transfer of energy from small to large angles over 
the course of the evolution of the jet, from highly relativistic to 
mildly relativistic regime, is also found to be small (Fig. \ref{fig1c}).

The Gaussian initial jet profile is a smooth and more realistic 
version of a `top hat' jet, where the hydrodynamic quantities are 
roughly constant within some typical opening angle, and sharply drop 
outside of this angle (though they drop smoothly, and not in a step 
function as in the `top hat' jet). We therefore expect the results
for an initial Gaussian profile to be similar to those of an initial
`top hat' jet profile. The hydrodynamic evolution of the latter has
been investigated using a 2D hydrodynamic simulation (Granot et al. 2001)
and was found to be quite similar to our results, namely the lateral 
spreading of the jet was much smaller than the prediction of simple 
`top hat' jet models, and there was very little lateral transfer of energy.
The fact that our hydrodynamic results for an initial Gaussian profile 
are similar to the results of 2D hydrodynamic simulations of an initial
`top hat' jet profile, is very reassuring and gives us some confidence 
in our numerical scheme.

\section{\large Light-curves}
\label{LC}

Once the jet dynamics and the pressure and density of the shocked
fluid are known, the synchrotron plus inverse-Compton emissions are
calculated from the fractional energies contained in the magnetic field 
and relativistic electrons which are parametrized by dimensionless numbers 
$\epsilon_B$ and $\epsilon_e$ respectively. The electrons are assumed to be 
accelerated to a power law distribution of energies, 
$dN/d\gamma_e\propto\gamma_e^{-p}$, promptly behind the shock, and then cool 
due to radiative losses and adiabatic cooling. The local emissivity,
$j'_{\nu'}$ (the energy emitted per unit volume per unit
time, frequency \& solid angle in the comoving frame), is approximated
by a broken power law, with breaks at the cooling frequency, $\nu'_c$,
the synchrotron frequency, $\nu'_m$, and the self absorption frequency, 
$\nu'_{sa}$.

Due to the curvature of the jet surface and its
motion, photons arriving at some observed time $t_{\rm obs}$
were emitted at different space-time points $({\bf r},t)$. 
The calculation of afterglow multi-wavelength light-curves takes
into account appropriate integration over equal arrival time surface
as given by the following equation for the flux density
\begin{equation}
\label{fnu}
F_\nu(t_{\rm obs},\hat{n})\delta t_{obs} = {(1+z)^2\over d_L^2} {d^4x 
   j'_{\nu'}({\bf r},t)
    \over \Gamma^2({\bf r},t) \left[ 1 - {\bf v\cdot\hat{n}}\right]^2},
\end{equation}
where prime denotes quantities in the comoving frame of the fluid, 
$\hat{n}$ is the direction to the observer (in the lab frame), 
$\nu'$ is related to the observed frequency $\nu$ by the Doppler shift 
relation i.e. $\nu'=\nu (1+z) \gamma (1 - {\bf v\cdot\hat{n}})$, 
$z$ and $d_L$ are the redshift and luminosity distance of the burst, 
and $d^4x$ is the Lorentz invariant 4-volume element. The observer time $t_{\rm obs}$ 
is related to the lab-frame time $t$ and location of the source {\bf r} by:  
$t_{\rm obs} = t - {\bf r}\cdot\hat{n}$, with 
${\bf r}\cdot\hat{n}=r(\cos\theta\cos\theta_{\rm obs}+\sin\theta\sin\theta_{\rm obs}\cos\phi)$.

\begin{figure}
\includegraphics[width=14.5cm]{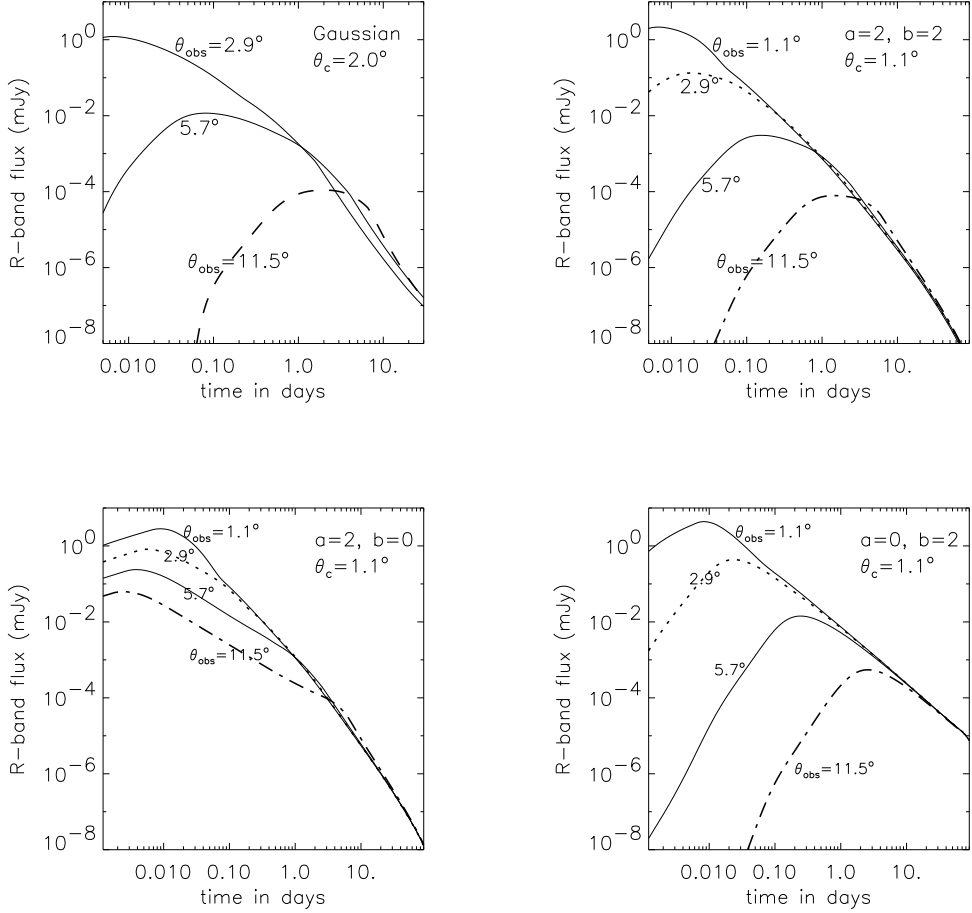}
\caption{\label{fig2}Observed R-band light-curves for different viewing angles, 
$\theta_{\rm obs}$, w.r.t. the jet axis are shown for the four jet profiles
presented in Figs. \ref{fig1a}-\ref{fig1d}; the basic jet model parameters are 
shown in the top right corner of each panel (see equation \ref{initial_con1} 
for the definition of $a$ \& $b$). 
The remaining model parameters are the same for all panels:
$z=1$, the energy fraction in electrons and magnetic field are
respectively $\epsilon_e = 0.5$ \& $\epsilon_B=10^{-4}$, the power-law 
index for electron distribution $p=2.5$, and a constant external density 
$n=10\;{\rm cm}^{-3}$. The lightcurves were calculated using the 
hydrodynamic simulation results shown in figures 1-4 (see appropriate 
figure captions for the details of the jet models) and equation 
(\ref{fnu}). For $a=b=2$ the temporal decay slope before the jet break 
is more moderate for large $\theta_{\rm obs}$, and if $\Gamma_0$ is not 
very large (200 in this figure) then the deceleration time for 
large $\theta_{\rm obs}$ is rather large. For $a=2$ and $b=0$ the
jet break is rather sharp (as seen in observations), but at
 $\theta_{\rm obs}\gtrsim$ a few $\theta_c$ there is a flattening
of the light curve just before the jet break (which is not seen 
in observations).
}
\end{figure}

The lightcurves for the four jet profiles described in \S \ref{NR} 
are shown in Figure \ref{fig2}, and the micro-physics parameters for the 
shocked gas can be found in the figure caption. The average thermal Lorentz
factor of shock heated protons is given by $\overline\Gamma_{th}(\theta,t) = 
1 + \hat\gamma\Pi(\theta,t)/\bigl[\mu_s(\theta,t)(\hat\gamma-1)\bigr]
 - \Pi/(\overline\Gamma^2\mu_s)$,
which turns out to be in very good agreement with the value one 
obtains from shock jump conditions.

The light curves for the Gaussian profile are rather similar to those for a
`top hat' jet,  although the jet break for `on axis'
observers ($\theta_{\rm obs}\lesssim\theta_c$) is somewhat less sharp for the 
Gaussian profile, compared to a `top hat' jet. The jet break is however still 
sufficiently sharp to be consistent with afterglow observations, in most cases.

For $a=0$ and $b=2$, the energy per unit solid angle ($\epsilon$) is 
independent of $\theta$, so that after the deceleration time, $t_{\rm dec}$, 
the light curves for viewing angles, $\theta_{\rm obs}<\theta_{\rm dec}$, 
become the same as for a spherical flow with energy $4\pi\epsilon$; note that
for this jet model $t_{\rm dec}\propto\Gamma^{-8/3}\propto
\theta_{\rm obs}^{16/3}$ and $\theta_{\rm dec}\propto t_{\rm obs}^{3/16}$.
These light curves do not show a jet break, and are therefore not compatible 
with afterglow lightcurves that have a break. 

For $a=2$ and $b=0$ (i.e. $\epsilon\propto\Theta^{-2}$ and $\Gamma=$const,
at initial time), there is a clear jet break for $\theta_{\rm obs}\gtrsim
2\theta_c$. The reason for this is that in this case much 
of the observed flux comes from small angles around the line of sight 
and the sharp break results when $\Gamma\lta\theta_{obs}^{-1}$.
For $\theta_{\rm obs}\gtrsim7\theta_c$ there 
is a flattening of the light curve just before the jet break, due to the 
contribution from the inner parts of the jet, which is not seen in the
observational data. This feature provides some constraint on this jet model
and is discussed in Granot \& Kumar (2002).

For $a=b=2$ (i.e. $\epsilon,~\Gamma\propto\Theta^{-2}$ initially) 
the temporal decay slope before the jet break is steeper for small 
$\theta_{\rm obs}$ and more moderate for large $\theta_{\rm obs}$,
and the magnitude of the increase in the slope after the steepning of
the lightcurve is larger for larger $\theta_{\rm obs}$. If such a correlation 
is found in the data, it would provide support for this jet profile. 
The jet break is smoother for small $\theta_{\rm obs}$,
and sharper for large $\theta_{\rm obs}$, making it difficult to explain the 
sharp jet breaks and small inferred $\theta_{\rm obs}$ (or opening angles for 
`top hat' jets) that have been observed in quite a few afterglows.

It can be seen that for $a=b=2$ and a small $\Gamma_0$ ($\sim$200 or less), 
the deceleration time, $t_{\rm dec}$, is quite large for large viewing
angles, so that the light curve has a rising part 
at $t<t_{\rm dec}$ (fig. 5). The fact that this is not seen in afterglow 
observations provides a lower limit on $\Gamma_0$ or an upper limits on $b$,
and thus can be used to constrain the structure of the jet. A more detailed 
analysis of the constraints that can be put on the jet profile from 
comparison to afterglow observations is discussed in an accompanying
paper (Granot \& Kumar 2002).
 
The light-curves obtained for the hydro simulation of
jets can be reproduced, quantitatively, by one or both
of two simple and extreme models, that are described below.
For more details on these models and their results we refer the reader 
to Granot \& Kumar (2002). The two different simple models are referred 
to as model 1 and model 2. 

In model 1, the energy per unit solid angle 
is assumed to retain its initial distribution, 
$\epsilon(\theta,t)=\epsilon(\theta,t_0)$. This represents the limiting 
case where there is very little lateral transport of energy. Model 2 attempts
to make the opposite assumption, that is, it assumes the maximal 
averaging of $\epsilon$ over the angle $\theta$ that 
is consistent with causality. The latter is achieved by averaging
$\epsilon$ over its initial distribution, over the range in $\theta$ out to
which a sound wave could have propagated from the initial time $t_0$.
These two extreme assumptions are designed to bracket the expected range 
of possible behaviors for lateral energy transport. They are therefore 
expected to roughly cover the range of observed flux which a more 
rigirous treatment of the jet dynamics should give. In this sense they 
serve to quantify the uncertainties in the jet dynamics and light curves.
For both models, the Lorentz factor is determined by energy conservation i.e.

\begin{equation}
\label{sim_dyna}
 \mu_0(\theta)\Gamma(\theta,t) + \mu_s(\theta,t)(\Gamma^2 -1) =
  \epsilon(\theta,t).
\end{equation}

\medskip
\section{Conclusion}
\label{conc}
\medskip

We have carried out hydrodynamical simulations of a relativistic, collimated,
axisymmetric outflow propagating into an external medium.
For simplicity, we used a uniform density medium for the calculations presented 
in this work. However, the numerical scheme developed in this paper is good 
for any axially symmetric external density distribution (including power laws 
with the distance from the source, etc.). We have reduced the problem to a 1-D 
system of partial differential equations by integrating over the radial thickness 
of the outflow, at a fixed lab-frame time, thereby greatly reducing the 
computation time. The hydrodynamical results were used to calculate the 
synchrotron emission and lightcurves for a variety of observer angles w.r.t. 
the symmetry axis. 

The model for GRB jets that is described in this paper is both rigorous and 
requires a very reasonable computational time, thus making it a useful tool
for the study of GRB afterglows. In particular, it can be used in fits to 
afterglow observations, which can help constrain the jet profile, the external 
density profile, and the micro-physics parameters of collisionless relativistic 
shocks.

We find that for jets with smoothly varying energy per unit solid angle
and Lorentz factor, $\Gamma$, the maximum transverse fluid velocity in 
the comoving frame of the shocked fluid is typically substantially less 
than the speed of sound: the peak velocity is of order 
$1/(\Gamma\,\delta\theta)$, where $\delta\theta$ is the angular scale 
for the variation of the energy per unit solid angle, or of $\Gamma$. 
Thus large transverse velocities, approaching the sound speed, 
are realized only when the energy density varies rapidly with $\theta$
or $\Gamma$ decreases to $(\delta\theta)^{-1}$. For the jet profiles 
examined in this paper, the largest lateral velocity occurred for the 
Gaussian profile, for which the initial gradients were largest. In fact, 
in this case a shock wave in the lateral direction developes due 
to the steep initial angular profile. This did not occur for the 
other jet profile which had a more moderate angular dependence. 

The energy per unit solid angle in the jet, $\epsilon$, is found to change
very slowly with time for all four of the jet models we have analyzed,
and to a reasonable approximation $\epsilon$ can be considered
essentially constant in time until $\Gamma$, along the jet axis, has dropped to
$\sim 4$, which corresponds to about a week since the explosion in the
observer frame (this is also the time when the transverse velocity
is getting to be of order $0.2\;c$). 

\begin{figure}
\includegraphics[width=16cm]{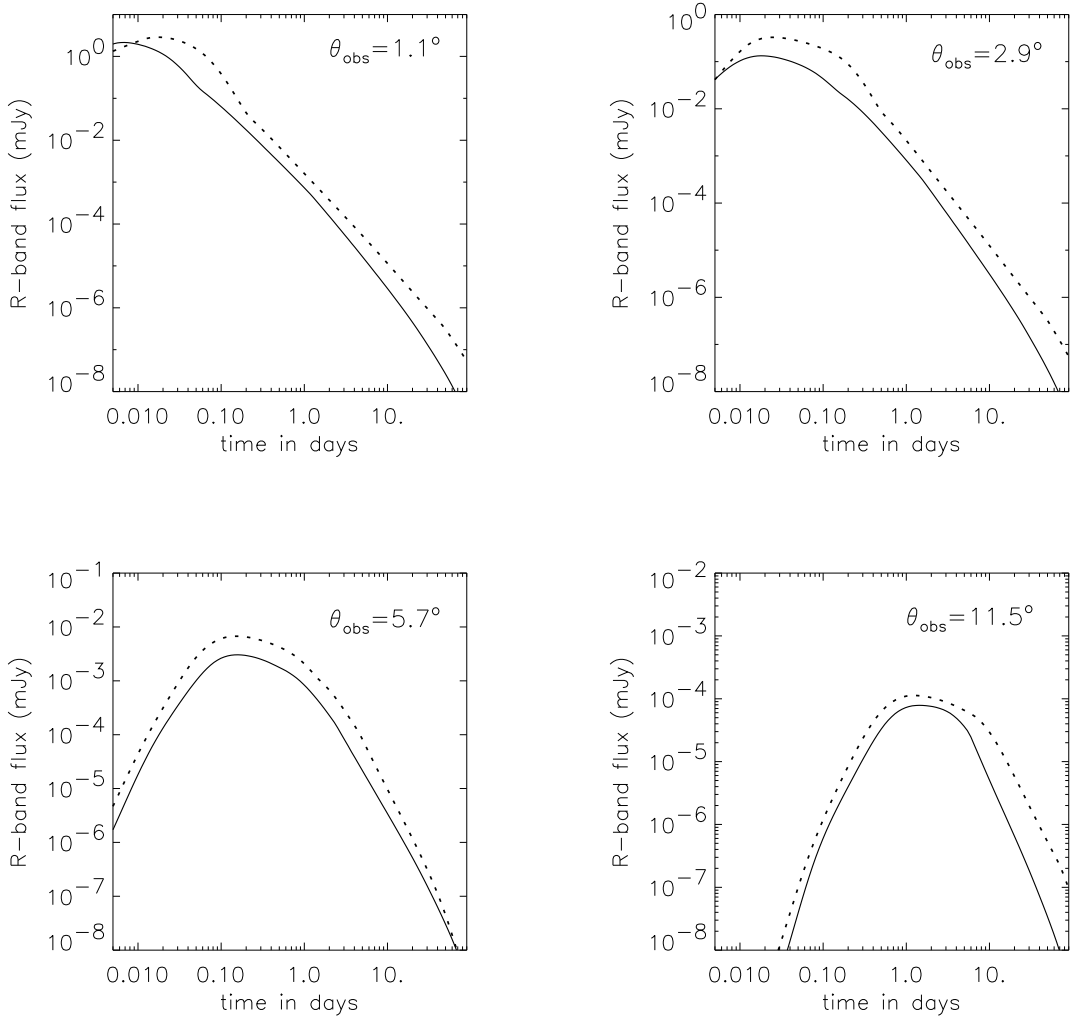}
\caption{\label{fig3}A comparison of lightcurves obtained with hydro 
simulation of jet (solid line) and a simple jet evolution model (dotted-curve) 
where the energy per unit solid
angle is assumed to be time independent. The initial jet model is a power-law
profile with $a=2$, $b=2$ and $\theta_c=1.1^o$ (see equation 
\ref{initial_con1} for the definition of $a$, $b$ \& $\theta_c$), 
and the observer location w.r.t. the jet axis, $\theta_{obs}$, is given 
in each panel.  All the other parameters for the calculation
are same as in figure 5 (see the caption for details).
For clarity the observed flux for the simple 
model has been multiplied by a factor 2.
}
\end{figure}

Therefore, a simple model where the energy per unit solid angle
is taken to be time independent, and each element of the jet behaves as if it 
were part of a spherical flow with the same $\epsilon$,
can serve as a useful approximation for the jet dynamics, 
as long as the jet is sufficiently relativistic. As can be seen from 
Figure \ref{fig3}, this simple model indeed seems to reproduce the 
light-curves obtained using the hydrodynamical simulations quite well. 
However, the simple model must necessarily breakdown when the transverse 
velocity becomes of order $c$, and the energy density 
is no longer constant. Unfortunately, this is also the regime when our 
hydrodynamical calculation becomes unstable.
 
We have calculated the observed light-curves in the R-band 
for several different jet angular profiles and viewing angles.
We find that the light curves for a Gaussian jet profile are similar
to those for a `top hat' jet (as expected), and compatible 
with most observed jet breaks. The light curves for a jet with a constant energy 
per unit solid angle, $\epsilon$, but $\Gamma(t_0)$ decreasing with 
$\theta$, are similar to those for a spherical explosion, and thus
not applicable to cases where we see a jet break in the light curve. 
For Jet profiles where initially $\epsilon\propto\Theta^{-2}$ and $\Gamma$ 
is either constant or $\propto\Theta^{-2}$ do produce jet breaks
in the light curves, and have the advantage that these models
can reproduce the `observed' narrow range for the total energy in GRB 
relativistic outflows (Panaitescu \& Kumar, 2002; Piran et al. 2001).
A companion paper (Granot \& Kumar, 2002) discusses some preliminary
constraints on jet profiles from qualitative comparison with observations.

\acknowledgements

JG thanks the support of the Institute for Advanced Study, 
funds for natural sciences.


\vfill\eject

\end{document}